\documentclass[aps,prl,groupedaddress,showpacs,showkeys,twocolumn]{revtex4}
\usepackage[]{natbib}
\usepackage{graphicx}
\usepackage{color}
\usepackage{dcolumn}
\usepackage{amsmath}
\usepackage{amssymb}
\usepackage[latin1]{inputenc}
\begin{document}

\title[]{Exploratory Behavior, Trap Models and Glass Transitions}

\author{\firstname{Alexandre} S. \surname{Martinez}}
\email{martinez@dfm.ffclrp.usp.br}

\author{\firstname{Osame} \surname{Kinouchi}}
\email{osame@dfm.ffclrp.usp.br}

\affiliation{Faculdade de Filosofia, Ci\^encias e Letras de Ribeir\~ao Preto, 
Universidade de S\~ao Paulo \\ 
Avenida Bandeirantes 3900, 
14040-901, Ribeir\~ao Preto, São Paulo, Brazil.}

 \author{\firstname{Sebastian} \surname{Risau-Gusman}}
 \email{srisau@if.ufrgs.br}
\affiliation{Instituto de F\'{\i}sica,
Universidade Federal do Rio Grande do Sul \\ 91501-970, Porto Alegre, Rio Grande do Sul, Brazil.}

\begin{abstract}
A random walk is performed on a disordered landscape composed of $N$ sites randomly and uniformly distributed inside a $d$-dimensional hypercube.
The walker hops from one site to another with probability proportional to $\exp [- \beta E(D)]$, where  $\beta = 1/T$ is the inverse of a formal temperature and $E(D)$ is an arbitrary cost function which depends on the hop distance $D$. 
Analytic results indicate that, if $E(D) = D^{d}$  and $N \rightarrow \infty$, there exists a glass transition at $\beta_d = \pi^{d/2}/\Gamma(d/2 + 1)$. 
Below $T_d$, the average trapping time diverges and the system falls into an out-of-equilibrium regime with aging phenomena.  
A L\'evy flight scenario and applications of exploratory behavior are considered.
\end{abstract}


\maketitle

This work focus on a surprising connection between two apparently disparated fields: models of exploratory behavior \cite{viswanathan2,viswanathan1} and the statistical physics of glass transitions described by trap models \cite{mezard:2002,bouchaud:1992,monthus:1996,bouchaud:2001}. 
We show that a glass transition may appear when a walker explores a random landscape 
with localized resources. 
This depends on a well defined manner of the cost function used to weight the possible movements. 
We have also obtained analytical results for finite size effects. 
This model is a simpler version of the stochastic tourist walk \cite{lima_prl2001,stanley_2001,kinouchi_phys_a,risaugusman:1:2003}.

The stochastic tourist walk can be viewed as a coarse-grained description of protein folding dynamics.
We represent low-lying minima by points in a $d$-dimensional configuration space. 
It is known that, in the glass phase, the configurational distance between minima constitute the dominant ``barrier'' (instead of energetic barriers) when considering coarse grained transitions between minima \cite{caliri:2003}. 
Another possible application of our results is in the study of charge diffusion over atoms which are random impurities in a substrate \cite{ferreira:2000}. 
An exponential decay of the transition probability as the configurational distance increases is a good approximation in both cases.

Trap models have been used to handle anomalous diffusion \cite{barkai:1998} and a low temperature regime in glass forming liquids where the system is supposed to hop among (free) energy minima \cite{dyre:1987,dyre:1995}.
In their simplest version \cite{bouchaud:1992,monthus:1996}, each state $i=1,\ldots,N$ represents a deep minimum which is endowed with a trap energy $E_i$ (``energy model'').  
When the system escapes from a given state, the model assumes that all other states have equal probability of being choosen as the next state. 
Alternatively, one can assume barriers $B_{ij}$ between the states $i$ and $j$ (``barrier model'') \cite{bouchaud:2001}. 
In such models, at some temperature $T_c$, there is a transition to an out-of-equilibrium regime  where the system average trapping time diverges and aging phenomena occurs.
This kind of scenario has been called ``weak ergodicity breaking'' transition \cite{bouchaud:2001,bertin:2002}.

Consider a random walk on a disordered landscape composed of $N$ sites (representing, for example, localized feeding sites as flowers, trees, water holes, islands etc.).
The coordinates of these sites are random and uniformly distributed along the unitary sides of $d$-dimensional hypercube. 
To perform a transition to another site, the walker uses a strategy based in some arbitrary cost function $E(D_{ij})$ which depends on the distance $D_{ij}$ between sites $i$ and $j$. 
Thus, $E(D_{ij})$ is similar to a barrier in trap models.
In our model the random variable is $D_{ij}$ with the probability distribution function (PDF) $P(D_{ij})$ fixed by the landscape. 
This contrasts to usual trap models, where $E$ is the random variable with $P(E)$ given by experimental evidences. 
Examples of the latter are Gumbel (exponential) distribution for extreme minima \cite{bertin:2002} and Gaussian PDF for super cooled liquids \cite{bouchaud:2002}.  
A further contrast is that  one has freedom to choose the cost function $E(D_{ij})$. 
Depending on its functional form, several scenarios can be envisaged for the decay of transition probabilities with  distance, for example exponential, Gaussian, power law etc. 
For each value of spatial dimension $d$, we have found the specific cost function that creates a glass transition which separates two distinct exploratory behaviors. 

The glass transition is characterized by a diverging mean residence time leading to out-of-equilibrium behavior. 
We have found analytically that the critical point is related to geometrical aspects of the distribution of sites and occurs at $\beta_{d} = 1/T_d = \pi^{d/2}/\Gamma(d/2 + 1)$, which is the volume of a $d$-dimensional hypersphere of unitary radius and $\Gamma(d/2 + 1)$ is the gamma function. 
Monte Carlo experiments validate the analytical analysis. 
We also show that the finite size and finite sample effects depend logarithmically on the number of sites and samples.

Consider a Poisson landscape where special (``feeding'') sites have coordinates $x_i^{(k)}$, with $i=1,2, \ldots, N$ and $k=1,2, \ldots,d$, that are uniform and randomly chosen in the interval $[0,1]$. 
A normalized Euclidean distance between sites $i$ and $j$:
$D_{ij}= N^{1/d} \{ \sum_{k=1}^d [ x^{(k)}_i-x^{(k)}_j]^2 \}^{1/2}$  
is introduced to simulate a constant density of points since the mean site separation is proportional to $N^{-1/d}$. 
The walker goes from site $i$ to any other site $j$, at each time step, with a ``thermal'' activation probability: 
\begin{equation}
W_{i \rightarrow j}  =  \frac{e^{-\beta E(D_{ij})}}{Z_i(\beta)} \; \; \; \mbox{with} \; \; \;Z_i(\beta) =  \sum_{j  = 1}^{N} e^{-\beta E(D_{ij})} \; , 
\end{equation} 
where $\beta$ is an external parameter which is the inverse of a formal temperature. 
The cost function $E(D_{ij})$   depends on the normalized distance $D_{ij}$ such that $E(D_{ii})=0$ allowing the walker remain in the site $i$.
Clearly, if $\beta \rightarrow \infty$, the walker remains in the present site ($W_{ii}=1$).
On the other hand, if $\beta \rightarrow 0$, the walker goes equally to any other site, regardless the distance.

Next we want to show that, although any site has a finite trapping time, the average residence time $\langle t_r \rangle$ diverges below a finite stochasticity level $T_d$. 
In this case, the system falls into a out-of-equilibrium regime and aging phenomena appears.

Consider discrete time steps.  
The probability of a walker to remain in a given site $a$ is $p_a(\beta)  \equiv  W_{a \rightarrow a} = 1/ Z_a$ and the probability that the walker leaves site $a$ is  $q_a(\beta) = 1 - p_a(\beta)$. 
Calculations become easier if the distances are relabeled according to the ordering with respect site $a$, so that the nearest neighbor of site $a$ is at $D_1^{(a)}$, the second nearest neighbor is at $D_2^{(a)}$ and so forth. 
One may write $Z_a(\beta) = 1 + \exp[- \beta E(D_1^{(a)})] \sum_{j = 1}^{N-1} \exp[- \beta \Delta_j^{(a)}]$ with: $\Delta_j^{(a)} =  E(D_j^{(a)}) - E(D_1^{(a)})$.

Given that at $t = 0$, the walker is at site $a$, the probability the walker remains there till time $t$ and leaves this site at $t + 1$ is given by the geometric distribution (the first failure after $t$ successes): $P_{\beta}(t) = p_a^t(\beta) q_a(\beta)$.
The residence (trapping) time associated to site $a$ is defined as the expected time: $t_r^{(a)}(\beta) =  \sum_{t = 0}^{\infty} t \; p_a^t(\beta) q_a(\beta) =  p_a(\beta)/[1 - p_a(\beta)]$, which reads:
$t_r^{(a)}(\beta)  = \exp[\beta E(D_1^{(a)})]/\{1 + \sum_{j = 2}^{N-1} \exp[- \beta \Delta_j^{(a)}]\}$.

The mean residence time for one realization of site distribution is: $t_r(\beta) = \sum_{a = 1}^{N} t_r^{(a)}(\beta)/N$ and the average over the disorder leads to:
\begin{eqnarray}
\nonumber
\langle t_r(\beta) \rangle                           & = &       \int \mbox{d} D_1 \ldots \mbox{d} D_{N-1} \; \frac{P(D_1, \ldots, D_{N-1}) \; t_1(\beta)}{1 + \sum_{j = 2}^{N-1} e^{- \beta \Delta_j}} \\ 
t_1(\beta)                 & \equiv & e^{\beta E(D_1)}  \label{eq:tr} \; .
\end{eqnarray}

To calculate the average of this trapping time over all possible sites, one needs to average over the full probability density distribution of neighbor distances $P(D_1, \ldots, D_{N-1})$.
This is a difficult task once it takes the boundaries into account. 
To remove the dependence on the boundaries one takes the thermodynamical limit $N \rightarrow \infty$.
Since the set of distances is an ordered set $D_1 < D_2 < \ldots < D_N$, the interdependence of the variables does not allow the factorization of the joint probability. 
One can make an approximation by considering the effect on the trapping time of the transitions to the first neighbor ($\Delta_j \rightarrow \infty$). 
Considering $\Delta_{N-1} > \ldots \Delta_2 > E(D_1)$, one can consider $\sum_{j = 2}^{N-1} \exp[- \beta \Delta_j] \ll 1$ and neglect this summation, which can also be viewed as a $\beta$ (high temperature) expansion. 
In this approximation one considers only the nearest neighbor distance PDF $P(D_1)$ leading to $\langle t_r(\beta) \rangle \cong  \langle t_1(\beta) \rangle$ where:
\begin{equation}
\label{eq:tr1}
\langle t_1(\beta) \rangle = \int_{0}^{\infty} \mbox{d} D_1 P(D_1) \: e^{\beta E(D_{1})} \; . 
\end{equation}
Considering bounds on the influence of the next neighbors
it is possible to show that $t_1(\beta)$ diverges, with the same exponent, at the same $\beta$ as the exact $t_r(\beta)$ as it be shown.

For a spatial Poisson process, one can easily show \cite{cox} that $P(D_1)  =  \exp[-V_d(D_1)] \mbox{d} V_d(D_1)/\mbox{d}D_1$, where $V_d(R)  =   A_d \; R^d$ is the volume of a hypersphere of radius $R$ and the geometrical factor (the volume of an unitary radius hypersphere) is $A_d = \pi^{d/2}/\Gamma(d/2 + 1)$, with $\Gamma(z)$ being the gamma function. 
Then: $P(D_1) = A_d \:d\: D_1^{d-1} e^{-V_d(D_1)}$ and
\begin{equation}
\label{eq:tr1b}
\langle t_1(\beta) \rangle =   A_d d \; \int_{0}^{\infty} \mbox{d} D_1 \; D_1^{d-1} \; e^{\left[\beta E(D_1) - A_d D_1^d \right] } \; .
\end{equation}

Firstly, consider a  cost function family which depends on a power of the distance, $E(D_{ij}) = D_{ij}^\alpha$, where $\alpha$ is an adjustable exponent.
For $\alpha<d$ the above integral is finite for any  $\beta$ value, so that, for long times, the system presents usual diffusive behavior.  
Conversely, for $\alpha > d$, the above integral is always divergent and the system is in an out-of-equilibrium regime. 
However, for $\alpha=d$, the residence time  and the tail of the distance distribution compete and a glass transition appears at finite value of $\beta$: $\langle t_1(\beta)  \rangle =   A_d  d   \int_{0}^{\infty}  \mbox{d}D_1 D_1^{d-1}\; e^{-D_1^d \left(A_d - \beta \right) } = A_d \int_{0}^{\infty} \mbox{d}x e^{-x \left(A_d - \beta\right) }$:
\begin{equation}
  \label{eq:t1_1} 
  \langle t_1(\beta)  \rangle  =  \left\{ \begin{array}{cl} A_d/(A_d - \beta) 
                             & (\beta < A_d) \\ \infty & (\beta \ge A_d) \end{array}\right. \; .
\end{equation}
The mean residence time is finite only for $\beta < A_d$ and it diverges at: 
\begin{equation}
\beta_{d} = A_d = \frac{\pi^{d/2}}{\Gamma(d/2+1)} \; , 
\end{equation}
which is just the volume of an unitary radius hypersphere in $d$ dimensions.
For example, $\beta_1 = 2$,  $\beta_2 = \pi$, $\beta_3 = 4 \pi /3$ etc. 
The value of $\beta_1$ has been also found in Ref. \cite{risaugusman:1:2003}, where the walker is not allowed to remain at the same site.

The residence time divergence at $\beta_d$ occurs only in the limit $N\rightarrow \infty$ and it is difficult to be observed in numerical experiments. 
Thus, finite size effects are important and they can be taken into account considering a cutoff value $D_c(N)$: $\langle t_1 \rangle  =  A_d \; d  \; \int_{0}^{D_c}  \mbox{d}D_1 \;  D_1^{d-1}\; e^{D_1^d \left(\beta - A_d \right) }$, and Eq. (\ref{eq:tr}) becomes:
\begin{equation}
\langle t_1 \rangle = \frac{A_d}{A_d-\beta} \left[ 1 - e^{-(A_d - \beta) D_c^d}\right]
                    \label{finite} \; .
\end{equation}
Notice that at $\beta = \beta_d$, the mean residence time is the volume of the largest hypersphere found in the $N_r$ realizations: $\langle t_1 \rangle = V_d(D_c) = A_d D_c^d$.

\begin{figure}[ht]
\begin{center}
\includegraphics[width=8.5cm]{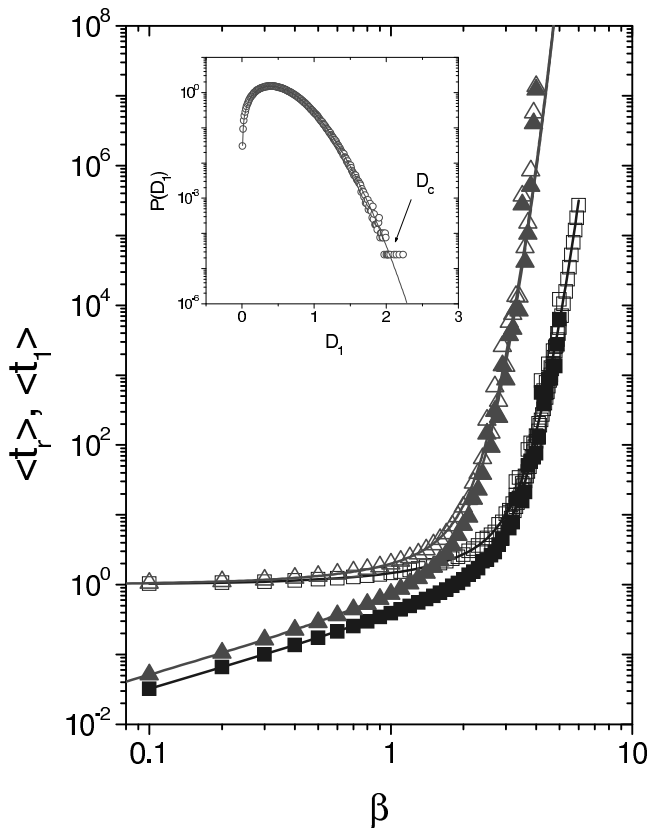}
\end{center}
\vspace{-0.5cm}
\caption{{\bf (a)} Monte Carlo calculations of $\langle t_1 \rangle$ for 
$d=1$ ($\triangle$) and $d=2$ ($\square$) 
and $\langle t_r\rangle$ ($\blacktriangle$, $\blacksquare$) 
which are compared to the curves Eqs.~(\ref{finite},\ref{eq:finite2}) (solid) as a function of $\beta$. 
Used values are: $N=100$ (with periodic boundary conditions for measuring distances), 
$M=10^4$ (that is, $N_r = 10^6$) leading to 
$D_c \approx 6.91$ for $d=1$ and $D_c \approx 2.10$ for $d=2$.
{\bf (b)} The same as before but on a log-log scale. {\bf Inset:} 
Histogram of sampled values of $D_1$ for $d = 2$ compared to $D_c = [  \ln ( N_r  )/\pi ]^{1/2}$.}
\label{fig:1}
\end{figure}

The value of $D_c$ can be estimated from extremal statistics concepts, being the most probable value of the Gumbel distribution \cite{mezard:1997}.
Consider a Monte Carlo calculation of Eq.~(\ref{eq:tr1}) where the average is taken over $M$ landscapes with $N$ points, i.e., $N_r = MN$ distances $D_1$ are sampled. 
The cut-off $D_c$ is obtained from the condition  $N_r \int_{D_c}^{\infty} \mbox{d}D_1 P(D_1) = 1$ which reads $N_r \exp[-V_d(D_c)] = 1$ or $D_c = [\ln N_r/A_d]^{1/d}$. 
For $d=1$, one has $D_c = \ln (N_r)/2$ and for $d=2$, $D_c = [  \ln ( N_r  )/\pi ]^{1/2}$.
In Fig.~\ref{fig:1} we compare Eq.~(\ref{finite}) with results from Monte Carlo calculations for $\langle t_1 \rangle$ and for the full residence time $\langle t_r\rangle$. 
Here we stress that in $\langle t_r\rangle$ the normalization $Z_i$ is calculated with all the terms $\exp[-\beta E(D_k)]$. 
Notice that, for $\beta=0$, we have $\langle t_1\rangle =1$ and $\langle t_r \rangle = 1/(N-1)$ but the two curves grows in the same way for larger $\beta$.

One can find the residence time PDF $P_T(t_1)$ at temperature $T$, in the first neighbor approximation.
Changing variables $ D_1 = (T \ln t_1)^{1/d}$ [see Eq.~(\ref{eq:tr})] in $P_T(t_1) = P[D_1(t_1)] \mbox{d}D_1/\mbox{d}t_1$ leads to $P_T(t_1)  =  A_d \:T \:t_1^{-(1+A_d T)}$, which has a power law tail $t_1^{-\gamma}$ with exponent $\gamma= 1+ A_d T$. 
As it is well known, the first moment diverges when the exponent is bellow $\gamma_c = 2$, that is, $T_d= 1/A_d$.

For $d=2$ the distance distribution function $P(D_1)= \pi D_1 \exp(- \pi D_1^2)$ presents a Gaussian tail. 
In this case, a well-defined glass transition appears when the cost function takes the form $E(D_{ij}) = D_{ij}^2$.
Notice that no glass transition occurs when $E(D_{ij})= D_{ij}$, although very long residence times exist when $\beta \rightarrow \infty$.
This case $E(D_{ij}) = D_{ij}$ is mathematically equivalent to Bouchaud's trap model for barrier distribution with Gaussian tail (with notation $B_{ij} = D_{ij}$). 
Thus, no glass transition is present at finite $T$ in the latter case \cite{bouchaud:2002}. 
To have a glass transition at finite $T$, the hopping probability in a Gaussian Bouchaud's trap model should have a Gaussian form $W_{i\rightarrow j} \propto \exp(- \beta B_{ij}^2)$. 

Now we consider an improved calculation for the mean residence time $\langle t_r(\beta) \rangle$ when $E(D) = N D^d$. 
The approximation consists to replace the distances to their mean values  \cite{percus:1996,chakraborti:2001} in the denominator of Eq. \ref{eq:tr}: $D_j = \langle D_j \rangle = (A_d N)^{-1/d} \Gamma(j + 1/d)/\Gamma(j) = [j/(N A_d)]^{1/d}$, where the approximation \cite{percus:1996}: $\Gamma(j + 1/d)/\Gamma(j) = j^{1/d}$ has been employed. 
In this way we obtain:  
\begin{equation}
\label{eq:finite2}
\langle t_{r}(\beta) \rangle =  \frac{1 - e^{- \beta/A_d}}{1 - e^{- (N-1)\beta/A_d}} \; \langle t_1(\beta) \rangle \; ,
\end{equation}
where $\langle t_1(\beta) \rangle$ is given by Eq. (\ref{eq:tr1}). 
As it can be seen in Fig. \ref{fig:1} this curve is in very good agreement with the Monte Carlo calculation.

The specific distribution $P(D_1) = \exp[-V_d(D_1)]$ studied here arises from the particular distribution of sites used (Poisson process), nevertheless other scenarios can be envisaged.
Consider, for example, a distribution of points where the distance PDF has a power law tail, $P(D_1) \propto D_1^{-b}$ (L\'evy process) \cite{mantegna_1994,koponen_1995,tsallis_levy}. 
For any cost function of the form $E= D_{ij}^\alpha$ the walker is always in the glass regime. 
In the integral of Eq.~(\ref{eq:tr1}), $P(D_1)$ cannot compete with an exponentially increasing residence times  $\exp(\beta D_1^\alpha)$. 
However, if the walker now uses a cost function of the form $E(D_{ij}) = \ln (1+D_{ij})$, both the transition probability and the residence time have power law forms, $W_{i \rightarrow j} = (1+D_{ij})^{-\beta}/Z_i$ and $t_1 = (1+D_1)^\beta$, respectively.
In this case, the average residence time is: $\langle t_1(\beta) \rangle \propto  \int \mbox{d}D_1 \; D_1^{(\beta-b)}$ and a glass transition occurs at $\beta_d = b-1$.

In a destructive foraging scenario \cite{viswanathan2} a walker wants to find unvisited feeding sites.
One can naively think that the walker should avoid glass transitions to reduce trapping times by choosing $\alpha < d$. 
Such values of $\alpha$ produce long hoppings which favor transitions to unvisited sites (the limit $\alpha \rightarrow 0$, $\beta \rightarrow \infty$ such that $\alpha \beta = \mbox{cte}$ produces L\'evy flights).
Nevertheless this procedure also increases the total traveling cost. 
We conjecture that the efficiency of this kind of exploration is maximized when the cost function allows a glass transition, for instance $E(D) = D^d$ in the Poisson case. 
Then, the optimal exploratory process may occur at some  temperature above the glass temperature $T_d$. 
This will be fully discussed elsewhere. 

Finally, we compare the present to previous models of exploratory behavior.
If one prohibits the walker to remain at the same site ($W_{a\rightarrow a} = 0$) the fundamental traps are cycles of period two instead of single sites. 
Also in this case we have found similar glass transitions \cite{risaugusman:1:2003}.
If one prohibits the walker to go to sites within a self-avoiding window of the $\tau$ past visited sites, the traps are cycles of diverse periods \cite{lima_prl2001,stanley_2001,kinouchi_phys_a}. 
In the latter case we also expect that the escaping from cycles is done through a glass transition which is a theme for a future study.

In conclusion, we have introduced a simple exploratory behavior model where the hop probability between feeding sites depends on the site distances and a formal temperature that controls the stochasticity level. 
We have presented a simple analytical treatment that predicts the existence of a glass transition. 
This transition depends on the competition between the walker cost function and the geometry of first neighbor distances. 
Our model brings glass transitions previously found in usual trap models to a very simple geometrical context. 
The approximations used give insight on the transition mechanism and accords well with results from numerical experiments.
Finite size effects depend logarithmically on the number of sites and sample for large $N$ and $M$.
The mean neighbor distance approximation on the calculation of the mean residence time extends the trap model approximation where only the first neighbor is considered.

\begin{acknowledgments}
The authors acknowledge useful conversations with N. Caticha, R. Dickman,  J. R. Drugowich de Felício, M. A. Idiart, G. F. Lima, R. da Silva and D. Stariolo. 
O. Kinouchi thanks financial support from FAPESP.
\end{acknowledgments}


\begin{thebibliography}{24}
\expandafter\ifx\csname natexlab\endcsname\relax\def\natexlab#1{#1}\fi
\expandafter\ifx\csname bibnamefont\endcsname\relax
  \def\bibnamefont#1{#1}\fi
\expandafter\ifx\csname bibfnamefont\endcsname\relax
  \def\bibfnamefont#1{#1}\fi
\expandafter\ifx\csname citenamefont\endcsname\relax
  \def\citenamefont#1{#1}\fi
\expandafter\ifx\csname url\endcsname\relax
  \def\url#1{\texttt{#1}}\fi
\expandafter\ifx\csname urlprefix\endcsname\relax\def\urlprefix{URL }\fi
\providecommand{\bibinfo}[2]{#2}
\providecommand{\eprint}[2][]{\url{#2}}

\bibitem[{\citenamefont{Viswanathan et~al.}(1999)\citenamefont{Viswanathan,
  Buldyrev, Havlin, da~Luz, Raposol, and Stanley}}]{viswanathan2}
\bibinfo{author}{\bibfnamefont{G.~M.} \bibnamefont{Viswanathan}},
  \bibinfo{author}{\bibfnamefont{S.}~\bibnamefont{Buldyrev}},
  \bibinfo{author}{\bibfnamefont{S.}~\bibnamefont{Havlin}},
  \bibinfo{author}{\bibfnamefont{M.~G.~E.} \bibnamefont{da~Luz}},
  \bibinfo{author}{\bibfnamefont{E.~P.} \bibnamefont{Raposol}},
  \bibnamefont{and} \bibinfo{author}{\bibfnamefont{H.~E.}
  \bibnamefont{Stanley}}, \bibinfo{journal}{Nature}
  \textbf{\bibinfo{volume}{401}}, \bibinfo{pages}{911} (\bibinfo{year}{1999}).

\bibitem[{\citenamefont{Viswanathan et~al.}(1996)\citenamefont{Viswanathan,
  Afanasyev, Buldyrev, Murphy, Prince, and Stanley}}]{viswanathan1}
\bibinfo{author}{\bibfnamefont{G.~M.} \bibnamefont{Viswanathan}},
  \bibinfo{author}{\bibfnamefont{V.}~\bibnamefont{Afanasyev}},
  \bibinfo{author}{\bibfnamefont{S.}~\bibnamefont{Buldyrev}},
  \bibinfo{author}{\bibfnamefont{E.~J.} \bibnamefont{Murphy}},
  \bibinfo{author}{\bibfnamefont{P.~A.} \bibnamefont{Prince}},
  \bibnamefont{and} \bibinfo{author}{\bibfnamefont{H.~E.}
  \bibnamefont{Stanley}}, \bibinfo{journal}{Nature}
  \textbf{\bibinfo{volume}{381}}, \bibinfo{pages}{413} (\bibinfo{year}{1996}).

\bibitem[{\citenamefont{M\'ezard}(2002)}]{mezard:2002}
\bibinfo{author}{\bibfnamefont{M.}~\bibnamefont{M\'ezard}},
  \bibinfo{journal}{Physica A} \textbf{\bibinfo{volume}{306}},
  \bibinfo{pages}{25} (\bibinfo{year}{2002}).

\bibitem[{\citenamefont{Bouchaud}(1992)}]{bouchaud:1992}
\bibinfo{author}{\bibfnamefont{J.-P.} \bibnamefont{Bouchaud}},
  \bibinfo{journal}{J. Phys. I France} \textbf{\bibinfo{volume}{2}},
  \bibinfo{pages}{1705} (\bibinfo{year}{1992}).

\bibitem[{\citenamefont{Monthus and Bouchaud}(1996)}]{monthus:1996}
\bibinfo{author}{\bibfnamefont{C.}~\bibnamefont{Monthus}} \bibnamefont{and}
  \bibinfo{author}{\bibfnamefont{J.-P.} \bibnamefont{Bouchaud}},
  \bibinfo{journal}{J. Phys. A} \textbf{\bibinfo{volume}{29}},
  \bibinfo{pages}{3847} (\bibinfo{year}{1996}).

\bibitem[{\citenamefont{Rinn et~al.}(2001)\citenamefont{Rinn, Maass, and
  Bouchaud}}]{bouchaud:2001}
\bibinfo{author}{\bibfnamefont{B.}~\bibnamefont{Rinn}},
  \bibinfo{author}{\bibfnamefont{P.}~\bibnamefont{Maass}}, \bibnamefont{and}
  \bibinfo{author}{\bibfnamefont{J.-P.} \bibnamefont{Bouchaud}},
  \bibinfo{journal}{Phys. Rev. B} \textbf{\bibinfo{volume}{64}},
  \bibinfo{pages}{art. no. 104417} (\bibinfo{year}{2001}).

\bibitem[{\citenamefont{Lima et~al.}(2001)\citenamefont{Lima, Martinez, and
  Kinouchi}}]{lima_prl2001}
\bibinfo{author}{\bibfnamefont{G.~F.} \bibnamefont{Lima}},
  \bibinfo{author}{\bibfnamefont{A.~S.} \bibnamefont{Martinez}},
  \bibnamefont{and} \bibinfo{author}{\bibfnamefont{O.}~\bibnamefont{Kinouchi}},
  \bibinfo{journal}{Phys. Rev. Lett.} \textbf{\bibinfo{volume}{87}},
  \bibinfo{pages}{010603} (\bibinfo{year}{2001}).

\bibitem[{\citenamefont{Stanley and Buldyrev}(2001)}]{stanley_2001}
\bibinfo{author}{\bibfnamefont{H.~E.} \bibnamefont{Stanley}} \bibnamefont{and}
  \bibinfo{author}{\bibfnamefont{S.~V.} \bibnamefont{Buldyrev}},
  \bibinfo{journal}{Nature} \textbf{\bibinfo{volume}{413}},
  \bibinfo{pages}{373} (\bibinfo{year}{2001}).

\bibitem[{\citenamefont{Kinouchi et~al.}(2002)\citenamefont{Kinouchi, Martinez,
  Lima, Louren\c{c}o, and Risau-Gusman}}]{kinouchi_phys_a}
\bibinfo{author}{\bibfnamefont{O.}~\bibnamefont{Kinouchi}},
  \bibinfo{author}{\bibfnamefont{A.~S.} \bibnamefont{Martinez}},
  \bibinfo{author}{\bibfnamefont{G.~F.} \bibnamefont{Lima}},
  \bibinfo{author}{\bibfnamefont{G.~M.} \bibnamefont{Louren\c{c}o}},
  \bibnamefont{and}
  \bibinfo{author}{\bibfnamefont{S.}~\bibnamefont{Risau-Gusman}},
  \bibinfo{journal}{Physica A} \textbf{\bibinfo{volume}{315}},
  \bibinfo{pages}{665} (\bibinfo{year}{2002}).

\bibitem[{\citenamefont{Risau-Gusman et~al.}(2003)\citenamefont{Risau-Gusman,
  Kinouchi, and Martinez}}]{risaugusman:1:2003}
\bibinfo{author}{\bibfnamefont{S.}~\bibnamefont{Risau-Gusman}},
  \bibinfo{author}{\bibfnamefont{O.}~\bibnamefont{Kinouchi}}, \bibnamefont{and}
  \bibinfo{author}{\bibfnamefont{A.~S.} \bibnamefont{Martinez}}
  (\bibinfo{year}{2003}), \bibinfo{note}{to appear in Phys. Rev. E, cond-mat/0301147}.

\bibitem[{\citenamefont{Barkay and Fleurov}(1998)}]{barkai:1998}
\bibinfo{author}{\bibfnamefont{E.}~\bibnamefont{Barkay}} \bibnamefont{and}
  \bibinfo{author}{\bibfnamefont{V.~N.} \bibnamefont{Fleurov}},
  \bibinfo{journal}{Phys. Rev. E} \textbf{\bibinfo{volume}{58}},
  \bibinfo{pages}{1296} (\bibinfo{year}{1998}).

\bibitem[{\citenamefont{Dyre}(1987)}]{dyre:1987}
\bibinfo{author}{\bibfnamefont{J.~C.} \bibnamefont{Dyre}},
  \bibinfo{journal}{Phys. Rev. Lett.} \textbf{\bibinfo{volume}{58}},
  \bibinfo{pages}{792} (\bibinfo{year}{1987}).

\bibitem[{\citenamefont{Dyre}(1995)}]{dyre:1995}
\bibinfo{author}{\bibfnamefont{J.~C.} \bibnamefont{Dyre}},
  \bibinfo{journal}{Phys. Rev. B} \textbf{\bibinfo{volume}{51}},
  \bibinfo{pages}{12276} (\bibinfo{year}{1995}).

\bibitem[{\citenamefont{Bertin and Bouchaud}(2002)}]{bertin:2002}
\bibinfo{author}{\bibfnamefont{E.~M.} \bibnamefont{Bertin}} \bibnamefont{and}
  \bibinfo{author}{\bibfnamefont{J.-P.} \bibnamefont{Bouchaud}}
  (\bibinfo{year}{2002}), \bibinfo{note}{cond-mat/0210521.}

\bibitem[{\citenamefont{Caliri}(2003)}]{caliri:2003}
\bibinfo{author}{\bibfnamefont{A.}~\bibnamefont{Caliri}}
  (\bibinfo{year}{2003}), \bibinfo{note}{private communication}.

\bibitem[{\citenamefont{Ferreira and Libero}(2000)}]{ferreira:2000}
\bibinfo{author}{\bibfnamefont{J.~V.~B.} \bibnamefont{Ferreira}}
  \bibnamefont{and} \bibinfo{author}{\bibfnamefont{V.~L.}
  \bibnamefont{Libero}}, \bibinfo{journal}{Phys. Rev. B}
  \textbf{\bibinfo{volume}{61}}, \bibinfo{pages}{10615} (\bibinfo{year}{2000}).

\bibitem[{\citenamefont{Denny et~al.}(2002)\citenamefont{Denny, Reichman., and
  Bouchaud}}]{bouchaud:2002}
\bibinfo{author}{\bibfnamefont{R.~A.} \bibnamefont{Denny}},
  \bibinfo{author}{\bibfnamefont{D.~R.} \bibnamefont{Reichman.}},
  \bibnamefont{and} \bibinfo{author}{\bibfnamefont{J.-P.}
  \bibnamefont{Bouchaud}} (\bibinfo{year}{2002}), \bibinfo{note}{cond-mat/0209020.}

\bibitem[{\citenamefont{Cox}(1981)}]{cox}
\bibinfo{author}{\bibfnamefont{T.~F.} \bibnamefont{Cox}},
  \bibinfo{journal}{Biometrics} \textbf{\bibinfo{volume}{37}},
  \bibinfo{pages}{367} (\bibinfo{year}{1981}).

\bibitem[{\citenamefont{Bouchaud and M\'ezard}(1997)}]{mezard:1997}
\bibinfo{author}{\bibfnamefont{J.-P.} \bibnamefont{Bouchaud}} \bibnamefont{and}
  \bibinfo{author}{\bibfnamefont{M.}~\bibnamefont{M\'ezard}},
  \bibinfo{journal}{J. Phys. A: Math. Gen.} \textbf{\bibinfo{volume}{30}},
  \bibinfo{pages}{7997} (\bibinfo{year}{1997}).

\bibitem[{\citenamefont{Percus and Martin}(1996)}]{percus:1996}
\bibinfo{author}{\bibfnamefont{A.~G.} \bibnamefont{Percus}} \bibnamefont{and}
  \bibinfo{author}{\bibfnamefont{O.~C.} \bibnamefont{Martin}},
  \bibinfo{journal}{Phys. Rev. Lett.} \textbf{\bibinfo{volume}{76}},
  \bibinfo{pages}{1188} (\bibinfo{year}{1996}).

\bibitem[{\citenamefont{Chakraborti}(2001)}]{chakraborti:2001}
\bibinfo{author}{\bibfnamefont{A.}~\bibnamefont{Chakraborti}},
  \bibinfo{journal}{Int. J. Mod. Phys. C} \textbf{\bibinfo{volume}{12}},
  \bibinfo{pages}{857} (\bibinfo{year}{2001}).

\bibitem[{\citenamefont{Mantegna and Stanley}(1994)}]{mantegna_1994}
\bibinfo{author}{\bibfnamefont{R.~N.} \bibnamefont{Mantegna}} \bibnamefont{and}
  \bibinfo{author}{\bibfnamefont{H.~E.} \bibnamefont{Stanley}},
  \bibinfo{journal}{Phys. Rev. Lett.} \textbf{\bibinfo{volume}{73}},
  \bibinfo{pages}{2946} (\bibinfo{year}{1994}).

\bibitem[{\citenamefont{Koponen}(1995)}]{koponen_1995}
\bibinfo{author}{\bibfnamefont{I.}~\bibnamefont{Koponen}},
  \bibinfo{journal}{Phys. Rev. E.} \textbf{\bibinfo{volume}{52}},
  \bibinfo{pages}{1197} (\bibinfo{year}{1995}).

\bibitem[{\citenamefont{Tsallis et~al.}(1995)\citenamefont{Tsallis, Levy,
  Souza, and Maynard}}]{tsallis_levy}
\bibinfo{author}{\bibfnamefont{C.}~\bibnamefont{Tsallis}},
  \bibinfo{author}{\bibfnamefont{S.~V.~F.} \bibnamefont{Levy}},
  \bibinfo{author}{\bibfnamefont{A.~M.~C.} \bibnamefont{Souza}},
  \bibnamefont{and} \bibinfo{author}{\bibfnamefont{R.}~\bibnamefont{Maynard}},
  \bibinfo{journal}{Phys. Rev. Lett.} \textbf{\bibinfo{volume}{75}},
  \bibinfo{pages}{3589} (\bibinfo{year}{1995}), \bibinfo{note}{{E}rratum: Phys.
  Rev. Lett. {\bf 77}, 5442 (1996).}

\end{thebibliography}

\end{document}